\documentclass[preprints,article,accept,moreauthors,pdftex]{Definitions/mdpi} 

\firstpage{1} 
\pubvolume{xx}
\issuenum{1}
\articlenumber{5}
\pubyear{2020}
\copyrightyear{2020}
\history{Received: date; Accepted: date; Published: date}
\pdfoutput=1

\usepackage{amsmath} % AMS Math Package
\usepackage{amssymb}	% Math symbols such as \mathbb
\usepackage{tensind}
\tensordelimiter{`}
\whenindex{a}{\alpha}{} \whenindex{b}{\beta}{} \whenindex{g}{\gamma}{} \whenindex{d}{\delta}{} \whenindex{e}{\varepsilon}{} \whenindex{m}{\mu}{} \whenindex{n}{\nu}{} \whenindex{l}{\lambda}{} \whenindex{k}{\kappa}{} \whenindex{r}{\rho}{} \whenindex{s}{\sigma}{} \whenindex{t}{\tau}{}
\usepackage[]{units}
\usepackage[all]{hypcap}
\makeatletter % Need for anything that contains an @ command 
\DeclareMathAlphabet{\ds}{U}{bbold}{m}{n}
\renewcommand{\v}[1]{\ensuremath{\mbox{\boldmath$ #1 $}}} 
\newcommand{\pd}[2]{\frac{\partial #1}{\partial #2}} 
\let\baraccent=\= % rename builtin command \= to \baraccent
\renewcommand{\=}[1]{\stackrel{#1}{=}} % for putting numbers above =
\newcommand{\OR}{\quad \vee \quad} %and symbol
\renewcommand{\(}{\left (}
\renewcommand{\)}{\right  )}
\renewcommand{\[}{\left [}
\renewcommand{\]}{\right ]}
\newcommand{\<}{\left <}
\renewcommand{\>}{\right >}

\theoremstyle{definition}

\theoremstyle{remark}

\newtheorem{?}{\textbf{Question}}

\newcommand{\e}[1]{\mbox{e}^{#1}} %exponential
\renewcommand{\exp}[1]{\mbox{exp}\[#1\]} %exponential
\newcommand{\bigO}[1]{O\[#1\]}  %order of (big O notation)
\Title{Causal Geometry}

\Author{Pavel Chvykov $^{1,}$*\orcidA{} and Erik Hoel $^{2}$}

\AuthorNames{Pavel Chvykov and Erik Hoel}

\address{%
$^{1}$ \quad Physics of Living Systems, Massachusetts Institute of Technology, Cambridge, MA 02139, USA; pchvykov@mit.edu\\
$^{2}$ \quad Allen Discovery Center, Tufts University, Medford, MA 02155, USA; erik.hoel@tufts.edu}

\corres{Correspondence: pchvykov at gmail com}
\abstract{Information geometry has offered a way to formally study the efficacy of scientific models by quantifying the impact of model parameters on the predicted effects. However, there has been little formal investigation of causation in this framework, despite causal models being a fundamental part of science and explanation. Here we introduce causal geometry, which formalizes not only how outcomes are impacted by parameters, but also how the parameters of a model can be intervened upon. Therefore we introduce a geometric version of ``effective information''---a known measure of the informativeness of a causal relationship. We show that it is given by the matching between the space of effects and the space of interventions, in the form of their geometric congruence. Therefore, given a fixed intervention capability, an effective causal model is one that matches those interventions. This is a consequence of ``causal emergence,'' wherein macroscopic causal relationships may carry more information than ``fundamental'' microscopic ones. We thus argue that a coarse-grained model may, paradoxically, be more informative than the microscopic one, especially when it better matches the scale of accessible interventions---as we illustrate on toy examples.}

\begin{document}

\section{Introduction}

Many complex real-world phenomena admit surprisingly simple descriptions, from~the smallest of microphysics to aspects of economics~\cite{hoel2018agent, anderson1972more, sole1996complxEdgeChaos}. While this is not seen as entirely coincidental, the~precise reasons for this fortunate circumstance are not entirely understood~\cite{zenil2012compIrred, israeli2004predictCA}. Two complementary solutions may be sought: On the one hand, we may hypothesize that this is an objective property of nature, thereby looking for some mechanism common among complex systems that allows them to be well-described with only a few parameters~\cite{sole1996complxEdgeChaos, transtrum2015IGperspective}. On~the other, we may guess that it is a subjective property of our perception and~then try to formalize the process by which we find useful patterns in arbitrarily complex systems~\cite{box1976all, israeli2004predictCA, maiwald2016modelingBio}. In~the recent years, substantial progress has been made in developing both of these perspectives, grounded in information theory~\cite{mackay2003info&learning}. 

One compelling argument for the first hypothesis has been made from information geometry. The~approach starts by associating with any given model a particular ``model manifold,'' whose geometric properties can tell us whether and which simplifications can be helpful~\cite{transtrum2015IGperspective, amari2016IGbook}. It turns out that for many real-world models, this manifold is highly anisotropic, having a hierarchical hyper-ribbon structure~\cite{sethna2007IGsysBio}. This property, termed ``sloppiness,'' indicates that only a few of the many microscopic model parameters are actually important for the model predictions---thus allowing for model simplification~\cite{transtrum2014mbam, machta2018priors}. While it is not yet clear how general this property is, sloppiness was empirically illustrated in a number of biochemical and physical models and~argued for on some general \mbox{grounds~\cite{machta2013IGscience, raju2018sloppyRG, transtrum2015IGperspective}.} This way, sloppiness provides an explanation for how emergent simplicity may be an objective property of complex systems~themselves. 

The second perspective instead takes as its starting point the well-known aphorism that ``all models are wrong, but~some are useful''~\cite{box1976all}. One way to see this is as a rejection of the reductionist notion that a ``fundamental'' microscopic description is the ``correct'' one, while all emergent system properties are derivatives from it~\cite{pines2000ToE, anderson1972more}. Instead, if~no model is seen as inherently best, then we can only compare models by their efficacy in predicting and controlling a given system. This amounts to finding an effective causal description of the system---such that distinct interventions causally translate to unique effects.
% This agrees with conventional intuition in that, e.g.,~how we model a tree will depend on whether we are interested in photosynthesis, ecology, or logging. 
In this perspective, simple descriptions may arise in part from our efforts to match the system description to how we interact with it~\cite{mackay2003info&learning, maiwald2016modelingBio}. 
%In this perspective, simple descriptions may arise not from any property of the complex system itself, but from our efforts to match the system description to how we interact with it~\cite{mackay2003info&learning, maiwald2016modelingBio}. 

%causality is importnat but hard. 
%A very different perspective arises when we focus on the causal structure of systems, which necessarily ties to our subjective intervention capabilities~\cite{pearl2018book}. 
Despite the fundamental role causality plays in science and its applications, its formal study has been scarce and elusive, until~recently~\cite{pearl2009causality}. Especially in complex systems, even conceptually defining what it means for something to be the cause of an observed phenomenon may be surprisingly tricky~\cite{halpern2016actualCaus}, let alone empirically verifying this~\cite{sugihara2012causComplEcosystems, solvang2019causComplSys}. In~particular, the~way models are studied in sloppiness makes no distinction between parameters as causes of the measured data, or~merely as its phenomenological descriptors (e.g., slope of a linear fit) or statistical~correlates.

Consider for example some bacterial population, whose size $ y(t)=\e{\theta\,t} $ at some time $ t $ depends on its growth rate $ \theta $. We cannot say that $ \theta $ is the {cause} of a large population; it remains merely a descriptor of the exponential fit, until~we introduce an additional ingredient: ``intervention capabilities''---some (at least hypothetical) way in which we can control this parameter and observe its effects. We can formalize this by recasting our system as a Bayesian dependency network and~defining the $ do $-operator~\cite{pearl2009causality}: $ do(\theta=\theta_0) $, which isolates the effects of actively setting the parameter $ \theta $ to value $ \theta_0 $, regardless of any other confounding factors that might influence it (such as if $ \theta $ were also affected by $ t $). The~role of the $ do $-operator is to distinguish the effect of a given intervention from that of all other possible interventions (or a lack thereof), thereby allowing a rigorous counter-factual definition of causality. This allows reliably disentangling causal dependencies $ p(y | do(\theta)) $ from mere statistical ones $ p(y | \theta) $~\cite{albantakis2019causal}. See Appendix~\ref{app:ex_do} for a worked out example explicitly illustrating how this distinction~arises.

Building on this, Judea Pearl developed the ``causal calculus'' framework, providing a set of tools to reliably work with intricate causal structures~\cite{pearl2009causality}. Further combining causal calculus with information theory allowed rigorously quantifying the amount of {Effective Information} ($ EI $%are the italics necessary?
) in the causal structure of a given model~\cite{tononi2003measuring}. In~a phenomenon termed ``causal emergence,'' it was then pointed out that in some systems, a~dimension-reduced description (such as a coarse-graining) may paradoxically carry more information about causation than the full microscopic model. This is because dimension reduction can sometimes yield a substantial reduction in noise and degeneracy~\cite{hoel2013quantifying, hoel2017map}.

In this work, we extend the causal emergence framework to continuous systems and~show that in that context, it is naturally related to information geometry and sloppy models. This leads to a novel construction, which we term causal geometry, where finding the causally most informative model translates to a geometric matching between our intervention capabilities and the effects on system behaviors, both expressed as distance metrics on the model's parameter space. This framework captures precisely how the inherent properties of the system's behavior (its sloppiness structure) and~their relation to its use context (matching to intervention capabilities) both play a role in optimal model selection, thereby reconciling the two above perspectives. This comes up because, on~the one hand, intervention capabilities are never unlimited in their degree or fineness and perfection of control, and~on the other, because~the behaviors of open physical systems are never without noise. This helps to formalize how neither the ``simplest'' nor the most ``fundamental'' reductionist model may be universally seen as~preferable.

We take a moment here to further clarify how this work fits within the current literature context. Besides~sloppiness and causal emergence, there have been other approaches developed for optimal model selection~\cite{antoulas07balTrunc, huang2005lumping}. For~example, Bayesian inference allows choosing the optimal modeling level by maximizing the posterior likelihood over a pre-defined library of models~\cite{daniels2015BayesModels}. In~contrast, the~sloppiness approach lacks a scalar quantity to be optimized for the best model and~only provides a heuristic for reducing unnecessarily complex models~\cite{transtrum2015IGperspective}. To~mitigate this, in~\cite{machta2018priors}, the authors optimized mutual information to find the set of coarse-grained parameters on sloppy manifolds that could reasonably be constrained by some limited experimental observations, thus selecting the optimal modeling level for the available data. The~mathematical procedure thus carried out closely paralleled that developed earlier in the causal emergence for discrete systems~\cite{hoel2013quantifying, hoel2017map}. In~our work, we use this formal resemblance to understand the connection between these two distant fields: sloppy models and causal calculus. This allows giving a continuum formulation of causal emergence, as~well as a novel local measure of causal optimality. On~the other side, our work establishes the proper formal role of interventions and causality in sloppiness, potentially resolving a long-standing formal challenge around the non-covariance of metric eigenvalues~\cite{dufresne2016rigorSlopp} and~showing that not only the hyper-ribbon manifold structure, but~also its relation to intervention capabilities account for the emergence of simple~models.

In Section~\ref{sec:continEI}, we define the Effective Information ($ EI $) for continuous models, which captures the amount of information in the model's causal relationships. We illustrate it on a simple example (Section \ref{sec:dimm_switch}) and~show how restricting the set of allowed interventions may sometimes, surprisingly, make the causal model more informative. Section~\ref{sec:causGeom} then introduces causal geometry. Specifically, it relates the continuous $ EI $ to information geometry, introducing a local geometric measure of causal structure, and~providing a way to find the locally most effective model for a given set of intervention capabilities using the techniques of information geometry. We demonstrate our construction on another simple toy model in Section~\ref{sec:2exp}, showing how causal emergence can arise in our geometric formulation, subject to the given interventional and observational~capabilities.

\section{Effective Information in Continuous Systems}
\label{sec:continEI}

For the purposes of this work, we formalize a {causal model} as a set of input-output relations, or~more precisely, a~map from all possible interventions to the full description of all effects within the context of some system~\cite{hoel2017map}. While the set of all hypothetically possible interventions on a given physical system is enormous and impractical to consider (involving arbitrary manipulations of every subatomic particle), the~set of experimentally doable (or even considered) interventions for a given context always represent a much smaller bounded space $ \mathcal{X} $, which we refer to here as ``intervention capabilities.'' Similarly, while an intervention will lead to uncountable microscopic physical effects, the~space of specific effects of interest $ \mathcal{Y} $ is much smaller and~often happens to be closely related to the intervention capabilities. All causal models by definition use some such subset of possibilities, and it is common in the literature around causation to restrict the set of hypotheticals, or~counterfactuals, within~a causal model~\cite{pearl2018book}. In~this work, we will illustrate how finding the optimal causal model for a given system is about a matching between the system behavior and the intervention capabilities~considered.

As the focus of this paper is on continuous systems, we consider $ \mathcal{X} $ and $ \mathcal{Y} $ to be continuous spaces, with~points $ \v{x} \in \mathcal{X} $ and $ \v{y} \in \mathcal{Y} $. To~formally discuss the causal model of our system, we make use of the $ do(x) $ operator, as~per Judea Pearl's causal calculus~\cite{pearl2009causality}. This operator is defined for any doable intervention the experimenter is capable of either performing or modeling, allowing assessing its causal effects. This allows us to formally describe a causal model as a map:
\begin{align}
\v{x} \to p(\v{y} \,|\, do(\v{x}))
\end{align}
where $ p $ is the probability density over effect space $ \mathcal{Y} $ resulting from ``doing'' the intervention $ \v{x} $. Note that this is distinct from $ p(\v{y} \,|\, \v{x}) $ in that the $ do $ operator allows us to distinguish the correlation introduced by the causal relation $ \v{x} \to \v{y} $ from one due to a common cause $ \v{a} \to \{\v{x},\v{y}\} $.

The notion of causality is then formalized as a counterfactual: How does the effect of $ do(\v{x}) $ differ from the effect of doing anything else? This latter ``null effect'' may include interventions such as $ do(\neg \v{x}) $, but~it can also include all other possible interventions: it is thus formally described by averaging together the effects of all considered intervention capabilities $\mathcal{X}$, giving the total ``effect distribution'':
\begin{align}
E_D(\v{y}) = \<p(\v{y} \,|\, do(\v{x}))\>_{\v{x}\in \mathcal{X}}
\end{align}
This way, to know precisely which effects $ do(\v{x}) $ causes, we can compare $ E_D(\v{y}) $ to $ p(\v{y} \;|\; do(\v{x})) $. The~distinguishability between these distributions may be captured with the Kullback--Leibler divergence $ D_{KL}\left [p(\v{y} \,|\, do(\v{x})) \; \| \; E_D(\v{y})\right ] $, giving us the amount of information associated with the application of an individual $ do(\v{x}) $ intervention~\cite{hoel2013quantifying, balduzzi2011information}. 

Averaging over all accessible interventions gives the information of the system's entire causal structure, termed the total ``effective information'':
\begin{align}
EI = \<D_{KL}\left [p(\v{y} \,|\, do(\v{x})) \; \| \; E_D(\v{y})\right ] \>_{\v{x}\in \mathcal{X}} \label{eq:EIdef}
\end{align}

Discrete versions of this effective information have been explored in Boolean networks~\cite{hoel2013quantifying} and graphs~\cite{klein2020emergence}. Note that the definition of $ EI $ here is identical to the mutual information between the uniform distribution over interventions $ I_D(\v{x})=const $ and the resulting distribution over effects $E_D(\v{y})$, so that: $ EI = \mathcal{I}(I_D; E_D) $~\cite{ hoel2017map, tononi2003measuring}. 

We proceed to illustrate with a simple example how the \textit{EI} varies across families of simple physical systems. As~such, we show how it may be used to select the systems that are in some sense ``best controllable,'' in that they best associate unique effects with unique interventions~\cite{liu2011controllability}. Additionally, this example will help us illustrate how the \textit{EI} may sometimes allow us to identify a coarse-grained system description that is more informative than the full microscopic one---thus illustrating causal emergence~\cite{hoel2017map}.

\subsection{Toy Example: Dimmer~Switch} \label{sec:dimm_switch}

\begin{figure}[H]
%\widefigure
\centering
\includegraphics[width=0.9\textwidth]{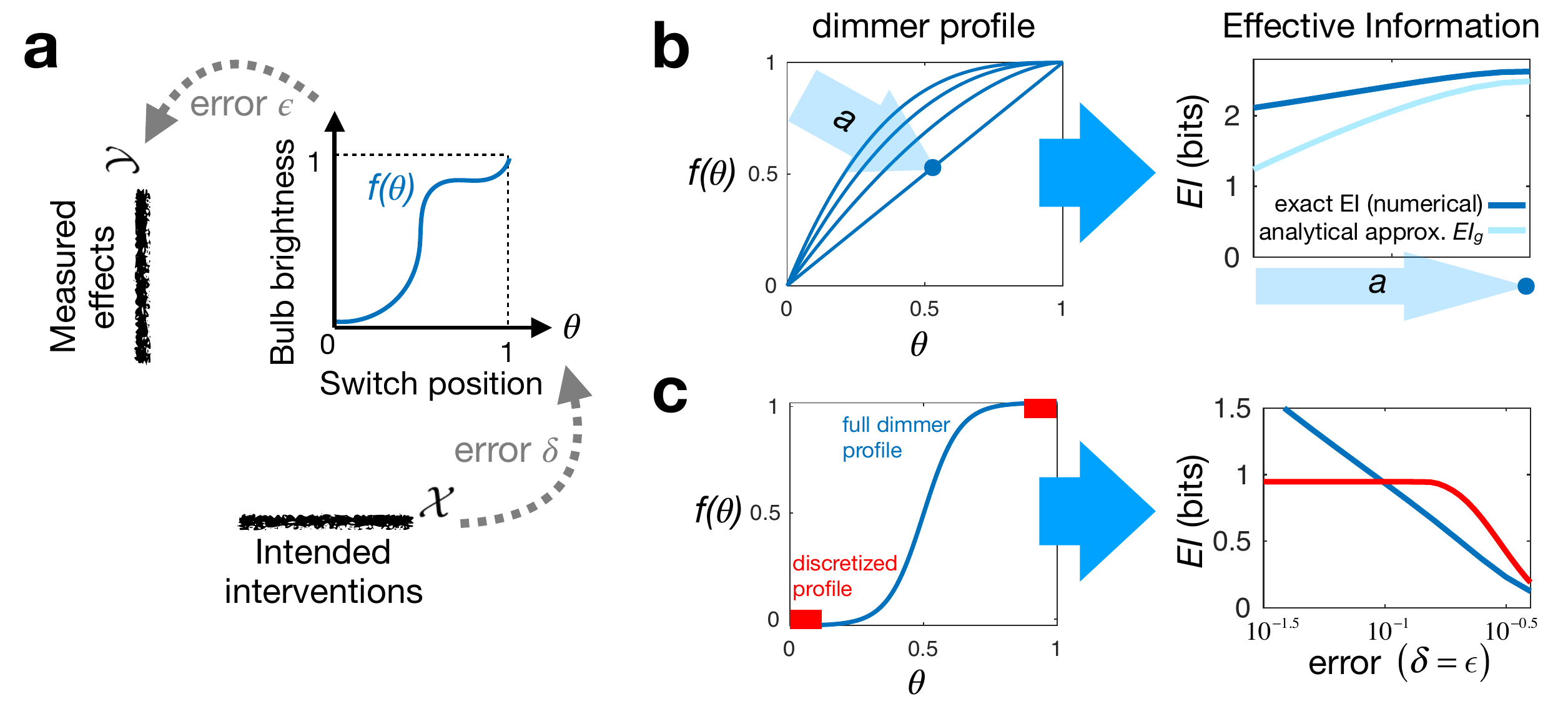}
\caption{
	Illustrating continuous Effective Information (\textit{EI}) on a simple toy system. (\textbf{a}) shows the system construction: a dimmer switch with a particular ``dimmer profile'' $ f(\theta) $. We can intervene on it by setting the switch $ \theta \in(0,1) $ up to error tolerance $ \delta $, while effects are similarly measured with error $ \epsilon $; (\textbf{b}) shows that for uniform errors $ \epsilon=\delta=0.03 $, out of the family of dimmer profiles parametrized by $ a $ (left), the~linear profile gives the ``best control,'' i.e.,~has the highest \textit{EI} (where dark blue---numerical \textit{EI} calculation and light blue---approximation in Equation~(\ref{eq:dimmer_EI})); (\textbf{c}) illustrates how for two other dimmer profiles (left), increasing error tolerances $ \epsilon=\delta $ influence the \textit{EI} (right, calculated numerically). The~profile in red represents a discrete binary switch---which emerges if we restrict the interventions on the blue dimmer profile to only use ``ends of run.'' Crucially, such coarse-graining allows for an {improved} control of the light (higher \textit{EI}) when errors are sufficiently large.
}
\label{fig:dimmer}
\end{figure}

Consider a continuous dimmer switch controlling a light bulb, but~with an arbitrary non-linear function $y=f(\theta)$, a~``dimmer profile,'' mapping from the switch setting $\theta \in \Theta=[0,1]$ to the light bulb brightness $y\in \mathcal{Y}=[0,1]$ (Figure\ref{fig:dimmer}a). To~quantify information about causation in continuous systems, we must carefully account for noise and errors in our inputs and outputs; else, infinite precision leads to infinite information. This is an issue for the application of all mutual information measures or their derivatives in deterministic continuous systems. Realistically, in~operating a dimmer switch, any user will have certain ``intervention error'' on setting its value, as~well as ``effect error,'' which can come either from intrinsic system noise or~from extrinsic measurement error. To~encode the effect error, we can replace the deterministic mapping $\theta \to y=f(\theta)$ with a probabilistic one $\theta \to p(y \,|\, do(\theta)) = \mathcal{N}_y(f(\theta),\epsilon^2)$---the normal distribution centered on $f(\theta)$ and with standard deviation $\epsilon$. While we could incorporate intervention error of setting $ \theta $ into this probability distribution as well, it is instructive for later discussion and generality to keep it separate.
The intervention error is thus similarly encoded by introducing a probabilistic mapping from the ``do%italics or not? please check the convention throughout
''-able interventions $x \in \mathcal{X}=[0,1]$ to the physical switch settings with some error $\delta$, as~ $x \to q(\theta\,|\, do(x))= \mathcal{N}_\theta(x,\delta^2)$. Here, we can think of the interventions $x \in \mathcal{X}$ as the ``intended'' switch settings, as~in practice, we cannot set the switch position with infinite precision. Note that while we do not explicitly model any possible confounding factors here, we assume that these may be present and important, but~are all taken care of by our use of the $ do $-operator. This ensures that only true causal relations, and~not spurious correlations, are captured by the distributions $ p $ and $ q $.
% start a new page without indent 4.6cm
%\clearpage
%\end{paracol}
%\nointerlineskip
%\begin{paracol}{2}
%\linenumbers
%\switchcolumn

With this setup, we can now use Equation~(\ref{eq:EIdef}) to explicitly compute an \textit{EI} for different dimmer profiles $f(\theta)$ and see which is causally most informative (has the most distinguishable effects). To~do this analytically for arbitrary $f(\theta)$, we must take the approximation that $\delta$ and $\epsilon$ are small compared to one (the range of interventions and effects) and~compared to the scale of curvature of $f(\theta)$ (such that $\epsilon\, f''(\theta) \ll f'(\theta)^2$). In~this limit, we have (for the derivation, see setup in Equation~(\ref{eq:CG_construct}) below and Appendix \ref{app:1D}):
\begin{align}
EI\simeq -\frac{1}{2}\; \int d \theta \; \log \left [2\pi e\(\(\frac{\epsilon}{f'(\theta)}\)^2 + \delta^2\)\right ],
\label{eq:dimmer_EI}
\end{align}
which echoes the form of the expression for entropy of a normal distribution. From~this, we see variationally that, given the fixed end-points $ f(0)=0 $ and $ f(1)=1 $, EI is maximized iff $ f'(\theta)=1 $: a uniformly linear dimmer switch. We can check this numerically by computing the exact EI for several different choices of $ f(\theta) $ (Figure \ref{fig:dimmer}b).

% For the straightforward scenario, this simple result is reasonable. 
A slightly more interesting version of this example is when our detector (eyes) perceives light brightness on a log, rather than linear, scale (Weber--Fechner law,~\cite{portugal2011weber-fechner}), in~which case the effect of the error will be non-uniform: $ \epsilon(y) \propto y$. If~this error is bound to be sufficiently small everywhere, Equation~(\ref{eq:dimmer_EI}) still holds, replacing only $ \epsilon \to \epsilon(y) \propto y=f(\theta) $. Again, we can variationally show that here, \textit{EI} is maximal iff $ f(\theta)/f'(\theta) $ is constant (up to fluctuations of magnitude $ \bigO{\delta/\epsilon} $), giving the optimal dimming profile $ f(\theta) = (\e{\theta /r} -1)/(\e{1/r}-1) $, with~some constant $ r \ll \delta/\epsilon $. In~reality, the~lighting industry produces switches with many dimming profiles that depend on the application~\cite{hu2016dimming}, so our approach can be seen as a principled way to optimize this~choice. 

Interestingly, restricting the accessible interventions can sometimes increase the amount of effective information if it increases the distinguishability, and~therefore informativeness, of~interventions. This is a form of causal emergence, wherein a higher level (coarsened) macroscopic model emerges as the more informative system description for modeling causation~\cite{hoel2017map}. To~give a particular example here, we compare the continuous dimmer profile shown in Figure~\ref{fig:dimmer}c (left, blue) to its discrete restriction (left, red)---which corresponds to a simple binary switch. 
%Note that while the hardware of the switch remains the same, and hence all settings $ \theta $ are still accessible, the change is in the user's space of intended interventions $ \mathcal{X} $. 
When the intervention and effect errors $ \delta =\epsilon $ are small, the~continuous switch gives more control opportunities and~is thus preferable; its \textit{EI} is larger than the 1 bit for the discrete switch. However, as~we increase the errors, we see a crossover in the two \textit{EI} values. In~this regime, the~errors are so large that the intermediate switch positions of the continuous profile become essentially useless and~are ``distracting'' from the more useful endpoint settings. Formally, such causal emergence arises due to the averaging over the set of all interventions in Equation~(\ref{eq:EIdef}). Practically, it captures the intuition that building good causal models, as~well as designing useful devices involve isolating only the most powerful control parameters out of all possible degrees of freedom~\cite{liu2011controllability}. 

\section{Causal~Geometry} \label{sec:causGeom}

Taking inspiration from information geometry, we can construct a more intuitive geometric expression for the \textit{EI}~\cite{machta2018priors}. For~studying causal structures of models, this ``geometric \textit{EI}'' we introduce may be viewed as a supplement to the usual \textit{EI} in Equation~(\ref{eq:EIdef}). While the geometric \textit{EI} corresponds to the \textit{EI} in a particular limit, we suggest that it remains useful more generally as an alternative metric of causal efficacy: it captures a causal model's informativeness like the \textit{EI} does, but~in a way that is local in a system's parameter space. In~this way, it frames causality not as a global counter-factual (comparing an intervention to all other interventions)~\cite{pearl2009causality}, but~as a local neighborhood counter-factual (comparing an intervention to nearby interventions). On~the flip side, our construct provides a novel formulation of information geometry that allows it to explicitly account for the causal relations of a model. Moreover, we argue that this is necessary for formal consistency when working with the Fisher information matrix eigenvalues (namely, their covariant formulation; see Section~\ref{sec:CG_IG})~\cite{dufresne2016rigorSlopp}. This suggests that model reduction based on these eigenvalues may not be made fully rigorous without explicitly accounting for causal relations in the~model.

\subsection{Construction}

Section~\ref{sec:continEI} described a causal model as a set of input-output relations between interventions $ \mathcal{X} $ and effects $ \mathcal{Y} $. Here, we investigate the relationship of such a causal model with the space of parameters $ \Theta $ that describe the underlying physical system. While these parameters need not necessarily have any direct physical meaning themselves, they are meant to give some abstract internal representation of the system---i.e., they mediate the mapping between interventions and effects. For~example, while the notion of energy is merely an abstract concept, it provides a useful model to mediate between interventions such as ``turning on the stove'' and effects like ``boiling water.'' The goal of our construction here is to compare how well different physical models capture the causal structure of a~system.

As we are focusing on continuous systems, we assume that our parameter space $ \Theta $ forms a smooth $ d $-dimensional manifold, with~parameters $ \theta_\mu $ indexed by $ \mu,\nu \in\{1,2,...,d\} $. Each accessible intervention $ \v{x}\in \mathcal{X} $ then maps to some probability distribution over parameters $ \v{x} \to q(\v{\theta}\, |\, do(\v{x})) $, and~each parameter in turn maps to a distribution over the observed effects $ \v{\theta} \to p(\v{y}\, |\, do(\v{\theta})) $. The~causal relations we are interested in here are thus simply $ \v{x} \to \v{\theta} \to \v{y} $, but~these are assumed to be embedded in some larger more complicated causal graph of additional confounding factors. These other possible hidden causes highlight the importance of using the $ do $-operator to isolate the causal relations in which we are interested (see Appendix~\ref{app:ex_do} for a simple example). Our one assumption is that the parameter space $ \Theta $ is chosen to be a sufficiently complete description of the system that no causal link can pass from $ \mathcal{X} $ to $ \mathcal{Y} $ directly without being reflected on $ \Theta $.

To understand the role of various parameters $ {\theta}_{\mu} $, we can ask how much the effects change as we perturb from some set of parameters in some direction: from $ \v{\theta} $ to $ \v{\theta}+\v{d\theta} $. Using the Kullback--Leibler divergence and~expanding it to leading order in $ \v{d\theta} $, we get:
\begin{align}
D_{KL}\left [p(\v{y} | do(\v{\theta})) \;\|\; p(\v{y} | do(\v{\theta}+\v{d\theta}))\right ] \simeq g_{\mu\nu}(\v{\theta}) \;d\theta_\mu \,d\theta_\nu 
\label{eq:FIM}
\end{align}
where summation over repeated indices is implied. This defines the Fisher information metric $ g_{\mu\nu}(\v{\theta}) = -\<\partial_\mu \partial_\nu \log p(\v{y} | do(\v{\theta}))\>_{p(\v{y} | \v{\theta})} $ with $ \partial_\mu \equiv \pd{}{\theta_\mu} $. This introduces a distance metric on the parameter space $ \Theta $, turning it into a Riemannian manifold, which we term the ``effect manifold'' $ \mathcal{M}_E $ (this is usually called simply the ``model manifold'' in the literature, but~here, we want to distinguish it from the ``intervention manifold,'' introduced below). More precisely, it is usually defined as $ \mathcal{M}_E \equiv \{p(\v{y}\, |\, do(\v{\theta}))\}_{\theta\in\Theta} $---the collection of all the effect distributions, or~the image of the parameter space $ \Theta $ under the model mapping, with~Equation~(\ref{eq:FIM}) being the natural distance metric on this space~\cite{amari2016IGbook, machta2013IGscience, transtrum2015IGperspective}.

Just as the mapping to effects defines the effect manifold $ \mathcal{M}_E $, we can similarly construct an ``intervention manifold'' $ \mathcal{M}_I $. For~this, we use Bayes' rule to invert the mapping from interventions to parameters $ \v{x} \to q(\v{\theta}\, |\, do(\v{x})) $, thus giving $ \v{\theta} \to \tilde{q}(do(\v{x}) \, |\, \v{\theta}) $---the probability that a given parameter point $ \v{\theta} $ was ``activated'' by an intervention $ \v{x} $. The~intervention manifold is thus defined as $ \mathcal{M}_I \equiv \{\tilde{q}(do(\v{x}) \, |\, \v{\theta})\}_{\theta \in \Theta} $, with~the corresponding Fisher information metric $ h_{\mu\nu} $ giving the distances on this space. With~this, we can now summarize our construction:

%MDPI: should it be the whole set of equations that is numbered by (6)? If yes, please put the (6) in the middle (as below). If no, please number each equation individually (remove the "/nonumber"s for each complete line)
%MDPI: ( if the latter, please be careful about which equation is labelled by \label{eq:CG_construct} ) %yes, it should be the whole set, thanks.

%\end{paracol}
%\nointerlineskip
\begin{equation} \label{eq:CG_construct}
\begin{split}
&\mbox{interventions }\v{x}\in\mathcal{X}, \mbox{ parameters } \v{\theta}\in\Theta \mbox{, effects } \v{y}\in\mathcal{Y}\\
&\v{x}\to q(\v{\theta}\,|\,do(\v{x})), \quad \v{\theta} \to p(\v{y}\,|\, do(\v{\theta}))\\
&\mbox{effect manifold }\mathcal{M}_E \equiv \{p(\v{y}\, |\, do(\v{\theta}))\}_{\theta\in\Theta} 
\quad \mbox{with metric } `g_mn`(\v{\theta}) = -\int d\v{y}\; p(\v{y}\, |\, \v{\theta})\; \partial_\mu \partial_\nu \ln p(\v{y}\, |\, do(\v{\theta})) \\%\quad \mbox{effect metric} \nonumber\\ 
&\mbox{intervention manifold } \mathcal{M}_I \equiv \{\tilde{q}(do(\v{x}) \, |\, \v{\theta})\}_{\theta \in \Theta}
\quad \mbox{with metric }
`h_mn`(\v{\theta}) = -\int d\v{x}\; \tilde{q}(do(\v{x})\, |\, \v{\theta})\; \partial_\mu \partial_\nu \ln \tilde{q}(\v{x}\, |\, \v{\theta}) \\ %\quad \mbox{intervention metric} \nonumber\\
& \mbox{where} \quad \tilde{q}(do(\v{x})\, |\, \v{\theta}) \equiv \frac{q(\v{\theta}\, |\, do(\v{x}))}{\int d\v{x}\; q(\v{\theta}\, |\, do(\v{x}))}
\quad \mbox{ and } \partial_\mu \equiv \pd{}{\theta_\mu} \mbox{ with } \mu,\nu \in \{1,2,...,d\}
\end{split}
\end{equation}

%\begin{paracol}{2}

%\linenumbers
%\switchcolumn

Note that for Bayesian inversion in the last line, we used a uniform prior over the intervention space $ I_D(\v{x}) = const $, which amounts to assuming that statistically, interventions are uniformly distributed over the entire considered space $ \mathcal{X} $. Note that this is not a choice of convenience, but~rather of conceptual necessity for correctly defining information in a causal model, as~argued in~\cite{hoel2017map}.
The natural point-wise correspondence between the two manifolds $ \mathcal{M}_E \leftrightarrow \mathcal{M}_I :\, p(\v{y} \, |\, do(\v{\theta}))\leftrightarrow \tilde{q}(do(\v{x}) \, |\, \v{\theta}) $ then allows for a local comparison between the two geometries. Alternatively, we may simply think of the parameter space $ \Theta $ with two separate distance metrics on it, effect metric $ g(\v{\theta}) $ and intervention metric $ h(\v{\theta}) $. With~this setup, we can now define our ``geometric'' effective information:
\begin{align}
& EI_{g} = \log\left [\frac{V_I}{(2\pi e)^{d/2}}\right ]-\< l(\v{\theta})\>_I \label{eq:EIg}\\
&\mbox{ with }\quad l(\v{\theta}) = \frac{1}{2}\log\, \det \(\ds{1} + g(\v{\theta})^{-1}\;h(\v{\theta})\) \label{eq:l}
\end{align}
Here, $ V_I $ is the volume of the intervention manifold $ \mathcal{M}_I $, which can be computed as $ V_I = \int d^d\v{\theta} \; \sqrt{\det h} $. It quantifies the effective number of distinct interventions we can $ do $, and~so, the first term in Equation~(\ref{eq:EIg}) gives the maximal possible amount of information about the causation our model could have, if~all interventions perfectly translated to effects. The~second term then discounts this number according to how poorly the interventions actually overlap with effects: geometrically, the~expression in Equation~(\ref{eq:l}) quantifies the degree of matching between the metrics $ g $ and $ h $ at the point $ \v{\theta} $ (here, $ \ds{1} $ stands for the identity matrix). This way, the~loss term $ l(\v{\theta}) $ can be interpreted as a measure of ``local mismatch'' between interventions and effects at $ \v{\theta} $, quantifying how much information about causation is lost by our modeling choice. The~average is then taken according to the intervention metric as: $ \<l(\v{\theta})\>_I \equiv \frac{1}{V_I}\int d^d\v{\theta} \; \sqrt{\det h} \;\;l(\v{\theta}) $. Note that the expression in Equation~(\ref{eq:EIg}) is identical to the approximation in Equation~(\ref{eq:dimmer_EI}) for the setup in that~example.

In Appendix \ref{app:multD}, we show that this expression for $ EI_g $ in Equation~(\ref{eq:EIg}) can be derived as the approximation of the exact $ EI $ in Equation~(\ref{eq:EIdef}) when both the mappings are close to deterministic: $p(\v{y}\,|\, do(\v{\theta})) = \mathcal{N}_{\v{y}}\(\v{f}(\v{\theta}), \epsilon^2\) $ and $ \tilde{q}(do(\v{x})\,|\, \v{\theta}) = \mathcal{N}_{\v{x}}\(\v{F}(\v{\theta}), \delta^2\) $, for~some functions $ \v{f}: \Theta\to \mathcal{Y} $ and $ \v{F}:\Theta\to \mathcal{X} $, with~small errors $ \epsilon $ and $ \delta $ (which may be anisotropic and nonuniform). Outside of this regime, the~$ EI $ and $ EI_g $ can differ. For~instance, while $ EI $ is positive by definition, $ EI_g $ can easily become negative, especially if $ g $ is degenerate anywhere on the manifold. Second, while $ EI $ captures the informativeness and therefore effectiveness of a causal model globally, $ EI_g $, and~more specifically the landscape $ l(\v{\theta}) $, can show us which local sectors of the parameter space are most and least causally effective. Finally, the~global nature of the computation for the exact $ EI $ quickly makes it intractable, even numerically, for~many continuous systems due to the proliferation of high-dimensional probability distributions---making $ EI_g $ the more practical choice in those~settings.

\subsection{Relation To~Sloppiness}\label{sec:CG_IG}

``Sloppiness'' is the property empirically observed in many real-world models, when the eigenvalues of the Fisher information matrix $ g_{\mu\nu} $ take on a hierarchy of vastly varying values~\cite{machta2013IGscience, sethna2007IGsysBio, transtrum2015IGperspective}. As~such, parameter variations in the directions corresponding to the smallest eigenvalues will have negligible impact on the effects~\cite{machta2013IGscience}. This leads to the hypothesis that we may effectively simplify our model by projecting out such directions, with~little loss for the model's descriptive power~\cite{transtrum2014mbam, machta2018priors}.

The trouble with this approach is that the components of the matrix $ g_{\mu\nu} $, and~hence its eigenvalues, depend on the particular choice of $ \theta $-coordinates on the effect manifold $ \mathcal{M}_E $~\cite{transtrum2015IGperspective, dufresne2016rigorSlopp}. Since the parameters $ \Theta $ represent some conceptual abstraction of the physical system, they constitute an arbitrary choice. This means that for a given point of $ \mathcal{M}_E $ labeled by $ \v{\theta} $, we can always choose some coordinates in which locally, $ g(\v{\theta}) = \ds{1}$ (an identity matrix), thus apparently breaking the above sloppiness structure. This issue is avoided in the literature by relying on the coordinate independent global properties of $ \mathcal{M}_E $, namely its boundary structure~\cite{transtrum2014mbam}.

Here, we show that by explicitly considering intervention capabilities, we can construct a local, but~still coordinate independent sloppiness metric. This becomes possible since interventions give a second independent distance metric on $ \Theta $~\cite{dufresne2016rigorSlopp}. The~matrix product $ g^{-1}h $ appearing in Equation~(\ref{eq:l}) is then a linear transformation, and~thus, its eigenvalues are coordinate independent. This way, to~evaluate how sloppy a given causal model is, we suggest that it is more appropriate to study the eigenvalues of $ h^{-1}g $ instead of those of $g$ as is usually done~\cite{transtrum2015IGperspective}. If~we then want to identify the directions in the parameter space that are locally least informative at a point $\v{\theta}$, we first need to re-express the metric $g$ in terms of the coordinates for which $h(\v{\theta}) = \ds{1}$ locally and~then find the appropriate eigenvectors in these new~coordinates. 

From this perspective, we see that the usual discussion of sloppiness, which does not study interventions explicitly~\cite{transtrum2015IGperspective}, may be said to implicitly assume that the intervention metric $ h(\v{\theta})\propto\ds{1} $, meaning that all model parameters directly correspond to physically doable interventions. Moreover, this requires that with respect to the given coordinate choice, each parameter can be intervened upon with equal uniform precision---which fixes the particular choice of coordinates on the parameter space. As~such, the~coordinate-specific eigenvalues $ \lambda_g(\v{\theta}) $ of the effect metric $ g(\v{\theta}) $ studied in information geometry, become physically meaningful in this special coordinate frame. In~particular, our expression for the local mismatch in Equation~(\ref{eq:l}) can here be expressed in terms of these eigenvalues as $ l(\v{\theta}) = \frac{1}{2} \sum_\lambda \log \(1+1/\lambda_g(\v{\theta})\)$. Thus, locally, the directions with the smallest $ \lambda_g $ account for the largest contribution to the mismatch $ l $. This recovers the standard intuition of sloppiness: we can best improve our model's descriptive efficacy by projecting out the smallest-$ \lambda_g $ directions~\cite{transtrum2014mbam, transtrum2011nonLinFits}. By~seeing how this result arises in our framework, we thus point out that it formally relies on the implicit assumption of uniform intervention capabilities over all model~parameters. 

It is worth noting that in the construction in Equation~(\ref{eq:CG_construct}), it may be possible to integrate out the parameters $ \v{\theta} $, giving directly the distribution of effects in terms of interventions $ P(\v{y}\, |\, do(\v{x})) = \int d^d\v{\theta}\; p(\v{y}\, |\, do(\v{\theta}))\, q(\v{\theta}\, |\, do(\v{x})) $ (we assume that $ \Theta $ gives a complete description of our system in the sense that no causal links from $ \mathcal{X} $ to $ \mathcal{Y} $ can bypass $ \Theta $). This way, the $ P(\v{y}\, |\, do(\v{x})) $ distribution gives an effect metric $ \hat{g}(\v{x}) $ over the $ \mathcal{X} $-space, which can directly quantify the amount of causal structure locally in our model. Nonetheless, the~intervention metric and Equation~(\ref{eq:l}) are still implicitly used here. This is because the space $ \mathcal{X} $ was constructed such that the intervention metric over it would be uniform $ \hat{h}(\v{x}) = \ds{1} $ everywhere. In~turn, Equation~(\ref{eq:l}) thus takes a particularly simple form in terms of $ \hat{g}(\v{x}) $. This illustrates that regardless of the parametrization we choose to describe our system, causal efficacy always arises as the matching between the effect metric and the intervention metric, per Equation~(\ref{eq:l})---which may or may not take on a simple form, and~must be checked explicitly in each case. Furthermore, though~this goes beyond the scope of this paper, we may imagine cases where the actual interventions form a complicated high-dimensional space $ \mathcal{X} $ that is harder to work with than the parameter space $ \Theta $ (just as the effect space $ \mathcal{Y} $ is often more complex than $ \Theta $). In~fact, this may be the typical scenario for real-world systems, where $ \mathcal{X} $ and $ \mathcal{Y} $ represent arbitrarily detailed descriptions of the system's context, while $ \Theta $ gives a manageable system~abstraction. 

\section{Two-Dimensional~Example} \label{sec:2exp}

In order to illustrate our causal geometry framework explicitly and show how higher level descriptions can emerge within it, we use a simple toy model (based on the example considered in~\cite{transtrum2014mbam, machta2018priors}). 

Imagine an experimenter has a mixed population of two non-interacting bacterial species that they are treating with two different antibiotics. The~experimenter's measurements cannot distinguish between the bacteria, and~so, they are monitoring only the total population size over time $ y(t) = \e{-\theta_1\,t} + \e{-\theta_2\,t}$, where $ \{\theta_1, \theta_2\} \in [0,1]$ are the death rates of the two individual species. These death rates are determined by the two antibiotic concentrations the experimenter treats the system with $ \{x_1,x_2\} $, which are the possible interventions here. In~the simplest case, each antibiotic will influence both species via some linear transformation $ A $, such that $ \theta_\mu = \sum_i A_{\mu i}\,x_i $. 

\begin{figure}[H]
%\widefigure
\centering
\includegraphics[width=0.9\textwidth]{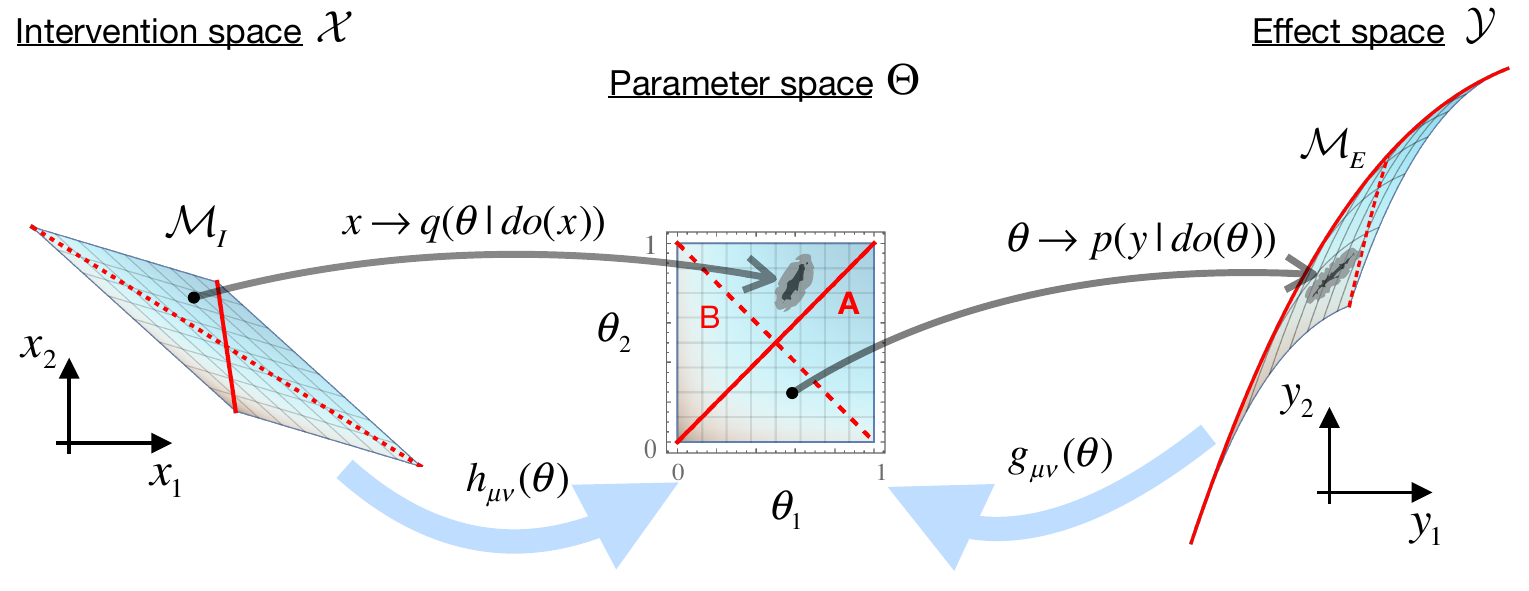}
\caption{An illustration of the causal geometry construction in Equation~(\ref{eq:CG_construct}). The~parameter space $ \Theta $ of our model gets two distinct geometric structures: the effect metric $ g_{\mu\nu}(\v{\theta}) $ and the intervention metric $ h_{\mu\nu}(\v{\theta}) $. Here, a~model is seen as a map that associates with each set of parameters $ \v{\theta} $, some distribution of possible measured effects $ \v{y} $ ({right}). As~parameters $ \v{\theta} $ may involve arbitrary abstractions and thus need not be directly controllable, we similarly associate them with practically doable interventions $ \v{x} $ ({left}). This way, our system description in terms of $ \v{\theta} $ ``mediates'' between the interventions and resulting effects in the causal~model.}
\label{fig:geom}
\end{figure}

%intention to look for CE using the submanifolds
This setup allows us to flesh-out the causal geometry construction and illustrate causal emergence here. Our main question is: When is this system best modeled microscopically, as~the two independent species, and~when does it behave more like a single homogeneous population, or~something else~\cite{tikhonov2016ecol}? To identify when higher scale models are more informative for the experimenter, we will calculate the geometric $ EI_g $ from Equation~(\ref{eq:EIg}) for the full 2D model described above and~then compare it to two separate 1D coarse-grained model descriptions, shown by the two red 1D sub-manifolds of the parameter space in Figure~\ref{fig:geom}.

\begin{figure}[H]
%\widefigure
\centering
\includegraphics[width=0.9\textwidth]{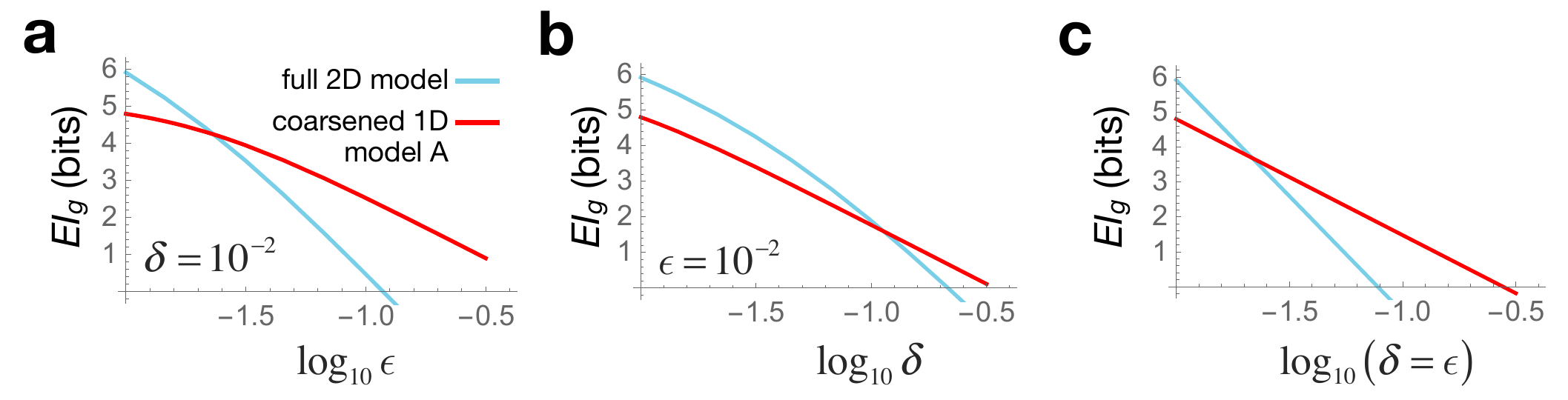}
\caption{Causal emergence from increasing errors for the toy model in Section~\ref{sec:2exp}. In~all panels, the blue line shows the $ EI_g $ for the full 2D model, while red for the 1D sub-manifold $ A $ shown in Figure~\ref{fig:geom} (solid red line). In~(\textbf{a}), we vary the effect error $ \epsilon $ at fixed intervention error $ \delta = 10^{-2} $; (\textbf{b}) varies intervention error $ \delta $ at fixed effect error $ \epsilon = 10^{-2} $; and~(\textbf{c}) varies both together $ \delta=\epsilon $. In~each case, we see a crossover where, with~no change in system behavior, the~coarse-grained 1D model becomes causally more informative when our intervention or effect errors become large.
}
\label{fig:2Enoise}
\end{figure}

We first specify the quantities for the construction in Equation~(\ref{eq:CG_construct}). Our interventions $ \v{x} $, having some uniform error tolerance $ \delta $, map to normal distributions over parameters $ \v{\theta} $ as: $ \v{x}\to q(\v{\theta}\,|\,do(\v{x})) = \mathcal{N}_{\theta}(A\v{x},\, A\,A^T \delta^2) $, giving the Bayesian inverse probability $ \tilde{q}(do(\v{x})\,|\,\v{\theta}) = \mathcal{N}_{x}(A^{-1}\v{\theta},\,\delta^2)$, and~hence the intervention metric $ h_{\mu\nu} = \sum_i (A^{-1})_{i\mu} (A^{-1})_{i\nu}/\delta^2$. The~effect space $ \v{y} $ is constructed by measuring the population size at several time-points, spaced out at intervals $ \Delta t $, such that the components of $ \v{y} $ are given by $ y_n = y(n\, \Delta t) = \e{-n \,\Delta t \,\theta_1} + \e{-n \,\Delta t \,\theta_2}$, with~$ n \in \{1,2,...,N\} $ and error $ \epsilon $ on each measurement (the initial conditions are thus always $ y(0)=2 $). Thus, we have $ \v{\theta} \to p(\v{y}\,|\, do(\v{\theta})) = \mathcal{N}_y\(\{y_n\}, \epsilon^2\)$ and~effect metric $ g_{\mu\nu} = \sum_n \partial_\mu y_n \, \partial_\nu y_n /\epsilon^2$. Figure~\ref{fig:geom} shows these mappings with $ N=2 $ for visual clarity, and~we use $ N=3 $ for the $ EI_g $ calculations below, but~all the qualitative behaviors remain the same for larger $ N $. Figure~\ref{fig:2Enoise} shows the resulting geometric $ EI_g $ (blue curves), computed via Equation~(\ref{eq:EIg}) for varying values of the error tolerances $ \epsilon $ and $ \delta $. 
% start a new page without indent 4.6cm

%\end{paracol}
%\nointerlineskip
%\clearpage

%\begin{paracol}{2}
%\linenumbers
%\switchcolumn

%explicit constructions for the calculation

% 1D submanifold, and CE in Figure~3
We can similarly find the $ EI_g $ for any sub-manifold of our parameter space, which would lead to a coarse-grained causal model, with~a correspondingly lower dimensional space of intervention capabilities. To~do this, we identify the pull-back of the two metrics in the full parameter space, to~the embedded sub-manifold, as~follows. We define a 1D submanifold of $ \Theta $ as a parametric curve $ (\theta_1,\theta_2) = (s_1(\sigma),s_2(\sigma))$ with the parameter $ \sigma $. The~pull-back effect metric on this 1D space with respect to $ \sigma $ will be the scalar $ \hat{g}(\sigma) = \sum_{\mu,\nu} s_\mu'(\sigma)s_\nu'(\sigma) \,g_{\mu\nu}(s_1(\sigma),s_2(\sigma)) $, and~similarly for intervention metric $ \hat{h}(\sigma) $. For~the 1D submanifold depicted by the solid red line in Figure~\ref{fig:geom}, the~resulting $ EI_g $ is plotted in red in Figure~\ref{fig:2Enoise}. 

The crossover seen in Figure~\ref{fig:2Enoise} thus illustrates causal emergence: for larger error values, the~coarse-grained 1D description turns out to be more informative than the full 2D model. Since this coarse-graining corresponds to the case where the two bacterial species are seen as identical $ \theta_1=\theta_2 $, we can say that at large errors, our bacterial colony is better modeled as a single homogeneous population. Crucially, this arises not from any change in system behavior, but~merely from how we interact with it: either from what interventions we impart or~from which effects we measure. Note also that when both the intervention and effect errors are scaled together $ \delta \propto \epsilon $, we see analytically from Equation~(\ref{eq:l}) that $ l(\theta) $ is constant, and~so:
\begin{align} \label{eq:crossover}
EI_g \sim \log V_I \sim -d\, \log\delta
\end{align}
which is also explicitly seen in Figure \ref{fig:2Enoise}c. %MDPI: please put this in a sentece: xxxxxxxxx, Figure \ref{fig:2Enoise}c. %done
This indicates that, quite generally, we expect to see crossovers between geometric $ EI $s of models with different $ d $ as we scale errors, with~low-dimensional models being preferred at large noise. Since noise is ubiquitous in all real-world complex systems, this argument suggests why reductionist microscopic descriptions are rarely optimal from the perspective of informative~interventions.

\begin{figure}[H]
	\centering
	\includegraphics[width=0.6\textwidth]{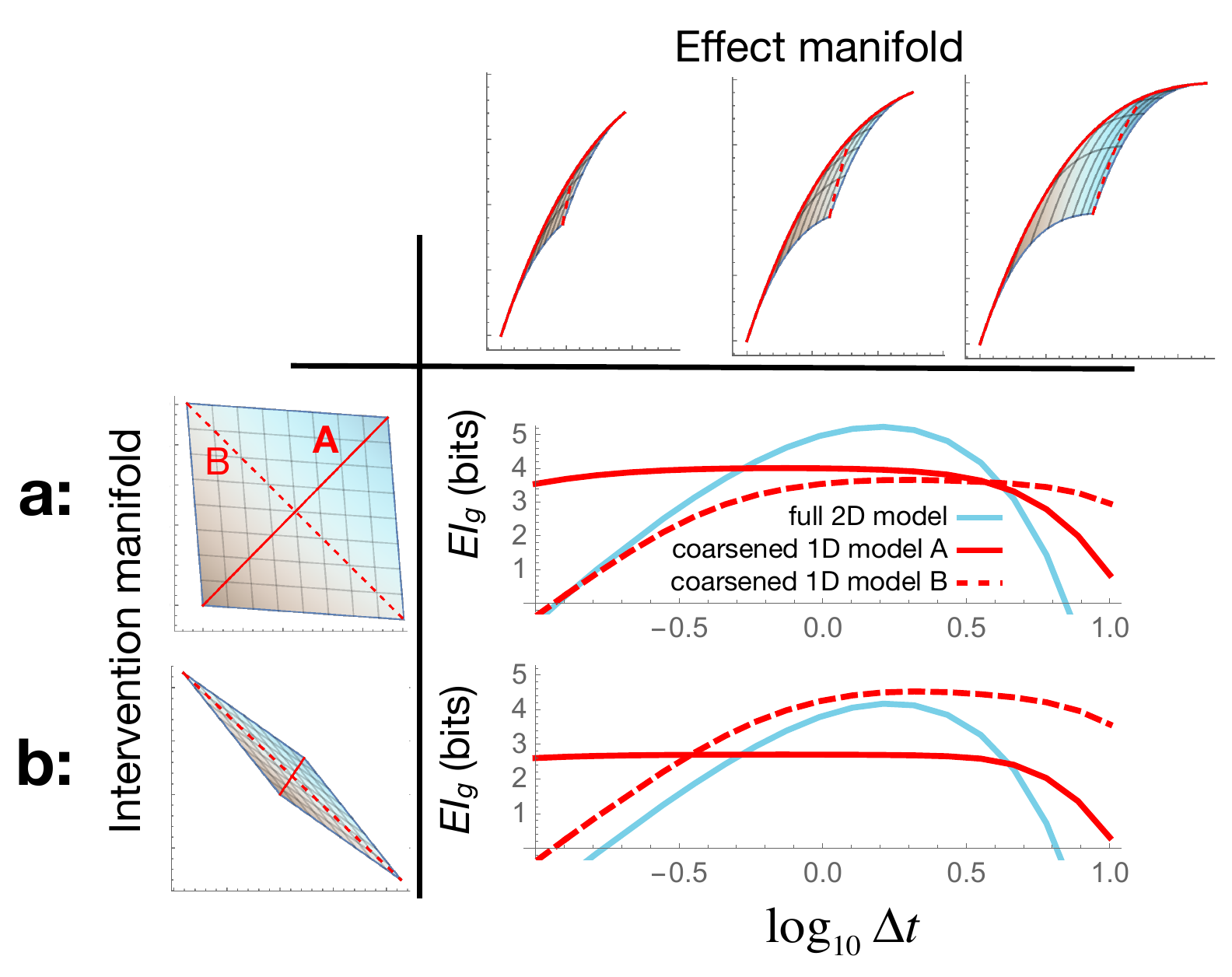}
	\caption{The optimal model choice depends on both the~effects we choose to measure and the intervention capabilities we have. Horizontally, we vary the time-scale $ \Delta t $ on which we measure the bacterial population dynamics in our toy model (Section \ref{sec:2exp}): the top row shows how this changes the shape of our effect manifold. (\textbf{a}) shows the results when our intervention capabilities are nearly in direct correspondence with the parameters $ \v{\theta} $. Here, the~$ EI_g $ plot shows that varying $ \Delta t $ takes us through three regimes: with submanifold $ A $ as the optimal model at early times, the~full 2D model optimal at intermediate times, and~submanifold $ B $ most informative at late times. (\textbf{b}) shows that this entire picture changes for a different set of intervention capabilities---illustrating that the appropriate model choice depends as much on the interventions as on the~effects. }
	\label{fig:2Einteff}
\end{figure}

By carrying out similar calculations, in~Figure~\ref{fig:2Einteff}, we explore how the optimal model choice depends on the time-scales we care about for the population dynamics (effects) and~the antibiotics we are using (interventions), all at fixed errors $ \epsilon, \,\delta $. 
When the two antibiotics control the two bacterial species almost independently ($ A \sim \ds{1} $, Figure~\ref{fig:2Einteff}a), we can identify three distinct regimes in the $ EI_g $ plot as we tune the measurement time-scale $ \Delta t $ along the $ x $-axis. If~we only care about the population's initial response to the treatment at early times, then we get a higher $ EI_g $ by modeling our colony as a single bacterial species. 
For intermediate times, the~full 2D model has the higher $ EI_g $, showing that in this regime, modeling both species independently is preferred. Finally, at~late times, most of the population is dead, and~the biggest remaining effect identifies how dissimilar their death rates were; the coarse-grained model given by the dashed red submanifold in Figure~\ref{fig:geom} ($ \theta_2 = 1 - \theta_1 $) turns out to be more informative. In~this regime, rather than viewing the population as either one or two independent bacterial species, we may think of it as a tightly coupled ecosystem of two competing species. Interestingly, such apparent coupling emerges here not from the underlying system dynamics, but~from the optimal choice of coarse-grained description for the given effects of~interest.

For a different set of intervention capabilities, where the antibiotics affect both species in a more interrelated way, with~$ A = \bigl( \begin{smallmatrix}1 & 0.8\\ 0.7 & 1\end{smallmatrix}\bigr) $, this entire picture changes (Figure \ref{fig:2Einteff}b). In~particular, we get the scenario where the ``fundamental'' two-species model is never useful, and~the unintuitive ``two competing species'' description is actually optimal at most time-scales. Note also that in all cases, for~very long and very short times $ \Delta t $, the~geometric $ EI_g $ drops below zero. While in this regime, the agreement with exact $ EI $ breaks down $ EI_g \neq EI $, it is also heuristically true that $ EI_g <0 \Rightarrow $ small $ EI $, and~so, the causal model is no longer very useful here. Even so, as~seen in Figure~\ref{fig:2Einteff}, some coarse-graining of the model may still be effective even when the full microscopic description becomes~useless. 

\section{Discussion}

The world appears to agents and experimenters as having a certain scale and boundaries. For~instance, solid objects are made of many loosely-connected atoms, yet in our everyday experience, we invariably view them as single units. While this may be intuitively understood in terms of the dictates of compression and memory storage~\cite{blumer1987occam}, this work proposes a way to formalize precisely how such a coarse-grained modeling choice may be the correct (causally optimal) one. This is particularly true for a given set of intervention capabilities. In~particular, we frame model selection as a geometric ``matching'' in the information space of the causal model to accessible~interventions. 

Intriguingly, this suggests that the correct choice of scientific modeling may not be merely a function of the correct understanding of the system, but~also of the context that system is being used in or the capabilities of the experimenters. Thus, for~example, if~the forces we used to handle solid objects were far larger than inter-atomic attraction holding them together, then viewing objects as single units would no longer be a good model. This echoes one of the main ideas in ``embodied cognition'' for AI and psychology, which posits that in order for an agent to build accurate models of reality, it needs the ability to actively intervene in the world, not merely observe it~\cite{shapiro2019embodied}.

This highlights a potentially important distinction between optimizing a model's predictive efficacy and its causal efficacy. Many approaches to optimal model selection, such as sloppiness, focus on getting computationally efficient predictions from a few fundamental parameters. In~contrast, optimizing causal efficacy looks for a model that best translates all interventions to unique effects, thus giving the user optimal power to control the system. Such a shift of motivation fundamentally changes our perspective on good scientific modeling: roughly, shifting the emphasis from prediction to control. While these two motivations may often go hand-in-hand, the~question of which is the more fundamental may be important in distinguishing~scenarios. 

%This highlights a potentially important distinction between optimizing a model's predictive efficacy and its causal efficacy. Many approaches to optimal model selection, such as sloppiness, focus on getting computationally efficient predictions from few fundamental parameters. In contrast, optimizing causal efficacy looks for a model that best translates all interventions to unique effects, thus giving the user optimal power to control the system. Such a shift of motivation fundamentally changes our perspective on good scientific modeling: roughly, shifting the emphasis from prediction to control. 
%%Rather than looking for a ``true'' system description, motivated by Occam's razor and predictive power, we look for a description of the world that gives us maximal ability to control it. 
%While these two motivations may often go hand-in-hand, the question of which is the more fundamental may be important in distinguishing scenarios. 

The causal geometry we introduce here is a natural extension of the information geometry framework~\cite{amari2016IGbook, transtrum2015IGperspective}, but~now explicitly accounting for the causal structure of model construction. In~our proposed formalism, a~given model becomes associated with two distinct Riemannian manifolds, along with a mapping between them, one capturing the role of interventions and~the other of the effects. The~relative geometric matching between these two manifolds locally tells us about how causally informative the present model is and~what coarse-graining may lead to a local~improvement. 

In this structure, the~colloquial notion of model ``sectors'' (especially used in field theories to refer to various field content~\cite{srednicki2007QFT}) becomes associated with literal sectors of the manifolds, with~their local geometries specifying the optimal or emergent descriptions of that sector. Such examples also highlight the importance of having a local way to quantify model optimality, as~globally, the manifold may have a complex and piecewise structure not amenable to simplification. While both traditional $ EI $~\cite{hoel2017map} and information-geometric model-reduction~\cite{transtrum2014mbam} depend on the global model behavior over the entire span of possible interventions, the~geometric $ EI_g $ introduced here is built by averaging inherently local causal efficacy, defined for each point in parameter space. We can further speculate that fundamentally, the~geometric matching in $ EI_g $ may provide a novel way to quantify causality locally, where the counter-factual comparison is considered relative to the local neighborhood of interventions, rather than to all globally accessible ones~\cite{pearl2009causality}.

We hope that causal geometry can contribute to further development in both formal principled methods for optimal model building in complex systems, as~well as an abstract understanding of what it means to develop informative scientific~theories. 

%%%%%%%%%%%%%%%%%%%%%%%%%%%%%%%%%%%%%%%%%%
\vspace{6pt} 

%%%%%%%%%%%%%%%%%%%%%%%%%%%%%%%%%%%%%%%%%%
%% optional
%\supplementary{The following are available online at \linksupplementary{s1}, Figure S1: title, Table S1: title, Video S1: title.}


%%%%%%%%%%%%%%%%%%%%%%%%%%%%%%%%%%%%%%%%%%
\authorcontributions{Conceptualization, P.C. and E.H.; formal analysis, P.C.; supervision, E.H.; visualization, P.C.; writing, original draft, P.C.; writing, review and editing, E.H. All authors read and agreed to the published version of the manuscript.}
%	Conceptualization, P.C. and E.H.; technical derivations and calculations, P.C.; writing, P.C. and E.H.. All authors have read and agreed to the published version of the manuscript.}

%%%%%%%%%%%%%%%%%%%%%%%%%%%%%%%%%%%%%%%%%%
\funding{E.H. was funded by Army Research Office Grant W911NF2010243 and~P.C. by ARO W911NF1810101 and~the James S. McDonnell Foundation Scholar Grant~220020476.}
%%%%%%%%%%%%%%%%%%%%%%%%%%%%%%%%%%%%%%%%%%

%\institutionalreview{Not applicable}
%MDPI: In this section, you should add the Institutional Review Board Statement and approval number, if relevant to your study. You might choose to exclude this statement if the study did not require ethical approval. Please note that the Editorial Office might ask you for further information. Please add ``The study was conducted according to the guidelines of the Declaration of Helsinki, and approved by the Institutional Review Board (or Ethics Committee) of NAME OF INSTITUTE (protocol code XXX and date of approval).'' OR ``Ethical review and approval were waived for this study, due to REASON (please provide a detailed justification).'' OR ``Not applicable'' for studies not involving humans or animals. %done

%\informedconsent{Not applicable}
%MDPI: Please add ``Informed consent was obtained from all subjects involved in the study.'' OR ``Patient consent was waived due to REASON (please provide a detailed justification).'' OR ``Not applicable'' for studies not involving humans.

%\dataavailability{The data presented in this study are available on request from the corresponding author. } 
%MDPI: Please refer to suggested Data Availability Statements in section “MDPI Research Data Policies” at \href{https://www.mdpi.com/ethics}{https://www.mdpi.com/ethics}. %The calculations were done in Mathematica notebooks, and some numerics in Matlab. It would not be too difficult to put together a presentable notebook with this if it is required by the journal (though it would take a few more days).

%%%%%%%%%%%%%%%%%%%%%%%%%%%%%%%%%%%%%%%%%%
\acknowledgments{Thanks Mark Transtrum for helpful discussions throughout the work on the manuscript, as~well as review of the final draft, Mikhail Tikhonov for inspiration at the early stages of the project, and Thomas Berrueta for great comments on the final~draft.}

%%%%%%%%%%%%%%%%%%%%%%%%%%%%%%%%%%%%%%%%%%
\conflictsofinterest{The authors declare no conflict of~interest.}
%	Declare conflicts of interest or state ``The authors declare no conflict of interest.'' Authors must identify and declare any personal circumstances or interest that may be perceived as inappropriately influencing the representation or interpretation of reported research results. Any role of the funders in the design of the study; in the collection, analyses or interpretation of data; in the writing of the manuscript, or in the decision to publish the results must be declared in this section. If there is no role, please state ``The funders had no role in the design of the study; in the collection, analyses, or interpretation of data; in the writing of the manuscript, or in the decision to publish the results''.} 

%%%%%%%%%%%%%%%%%%%%%%%%%%%%%%%%%%%%%%%%%%
%% optional
%\abbreviations{The following abbreviations are used in this manuscript:\\
%
%\noindent 
%\begin{tabular}{@{}ll}
%$ EI $ & Effective Information\\
%%CE & Causal Emergence\\
%%IG & Information Geometry\\
%\end{tabular}}

%%%%%%%%%%%%%%%%%%%%%%%%%%%%%%%%%%%%%%%%%%
%% optional
\appendixtitles{yes} % Leave argument ''no'' if all appendix headings stay EMPTY (then no dot is printed after ''Appendix A''). If~the appendix sections contain a heading then change the argument to ''yes''.
\appendix

\section{Example Illustrating the \textit{do}-operator} \label{app:ex_do}
Here, we give a simple pedagogic example that illustrates the use of the \textit{do}-operator and its role in distinguishing causal relations from correlation. To~highlight its role in the context of information geometry and sloppiness, we use a setup familiar from that literature~\cite{transtrum2015IGperspective}. Consider some bacterial population whose decay over time in unfavorable conditions we want to monitor: $ y(t) = \e{-\theta \, t} $, where $ y(t) $ is the remaining fraction at time $ t $ of the original population at $ t=0 $ and~$ \theta $ is the death rate. In~the context of information geometry, we might consider $ \theta $ as a model parameter, while $ y(t) $ as the predicted data. However, we cannot say that $ \theta $ ``causes'' the decay, as~any such statement about causality requires distinguishing interventions and effects. Without~this, our model is merely an exponential fit to the data $ y(t) $, with~$ \theta $ labeling a compressed representation of the decay~curve.

%To have a causal model, 
To have a causal, rather than a descriptive model, we need to introduce interventions that can influence this decay curve (and hence its descriptor $ \theta $). For~example, this may be the concentration of some harmful chemical in the bacteria's environment. For the~simplicity of our example, imagine that we have a direct linear mapping from these concentrations $ x $ to the death rate: $ do(x) \to \theta $. Note that this mapping is a causal model, with~the $ do $-operator being well-defined on the space of interventions: it prescribes actively setting the chemical concentrations, in~spite of any other possible environmental causes and fluctuations. 
% while the mathematical structure of our example did not change, we are now explicitly talking about a causal model, not a regression to the data. 
In this setting, we can show how causal dependencies $ q(\theta|do(x)) $ can differ from statistical ones $ q(\theta|x) $; these will be distinct whenever there are any confounding factors present. For~concreteness in our example, we can introduce temperature $ T $ as such a confounder: On the one hand, it might speed up bacteria's life-cycle, and~hence death rate such that $ do(x) \to \theta =x + \alpha\, T $. On~the other, it can denature the harmful chemical as $ x = x_0/T $, where $ x_0 $ is the reference concentration at $ T=1 $.

If our model is specified explicitly, as~is being done here, then all the causal relations are a priori known, and~no work is needed to distinguish causation from correlation. The~$ do $-operator becomes important when we look at a more real-world setting where we can interact with a system, but do not know its underlying dynamics. From~the perspective of sloppiness, this is referred to as data-driven manifold exploration. The~$ do $-operator prescribes how we should collect our data to extract the correct causal~relations. 

First consider a ``wrong'' way we can explore the model manifold, which will yield only statistical dependencies $ q(\theta|x) $. For~this, we study the fluctuations of bacterial population and habitat in their natural environment (specifically looking at $ \theta $ and $ x $). We can imagine that these fluctuations are dominated by natural variation of the reference chemical concentration $ x_0 \sim \mathcal{N}\(\hat{x}, \sigma_x^2\)$, with~an additional influence from temperature fluctuations $ T\sim \mathcal{N}\(0,\sigma_T^2\) $ (where we defined the scale of $ T $ such that its fluctuations centered on zero). Since we only observe $ x $, and~not $ x_0 $ or $ T $, we integrate these out to get:
\begin{align}
q(\theta|x) &= \frac{1}{Z(x)} \int dx_0 \; \e{-\frac{(x_0 - \hat{x})^2}{2\sigma_x^2}}\int dT \; \e{-\frac{T^2}{2\sigma_T^2}} \; \delta\(\theta - x - \alpha\, T\) \, \delta\(x_0 - x \,T\)\nonumber\\
&= \frac{1}{\alpha\, Z(x)} \;\e{-\frac{(x\, \frac{\theta-x}{\alpha} - \hat{x})^2}{2\sigma_x^2} - \frac{(\frac{\theta-x}{\alpha})^2}{2\sigma_T^2}}
= \frac{1}{\sqrt{2\pi \,\sigma_{net}^2}}\; \e{-\frac{1}{2 \sigma_{net}^2}\(\theta - x \(1+\frac{\hat{x}\, \sigma_{net}^2}{\alpha\, \sigma_x^2}\)\)^2 } \label{eq:statDep}
\end{align}
where we defined the $ x $-dependent $ \sigma_{net} \equiv \alpha \,\(\frac{1}{\sigma_T^2}+ \frac{x^2}{\sigma_x^2}\)^{-1/2} $ in the last line. 
%We thus see that the resulting distribution is also Gaussian in $ \theta $. 

Now, to~capture the correct causal relationships, we instead use the $ do $-operator to find $ q(\theta|do(x)) $, which prescribes actively setting the chemical concentration $ x $ to specific values, rather than passively waiting for these to be observed. This means that while temperature can still fluctuate as above $ T\sim \mathcal{N}\(0,\sigma_T^2\) $, we do not defer to fluctuations of $ x_0 $ to produce specific values of $ x $, but~rather set these concentrations ourselves (e.g., in~lab conditions). This removes confounding correlations, while $ T $ fluctuations now merely add uniform noise on $ \theta $:
% We can similarly integrate out $ T $ fluctuations here to get
\begin{align}
q(\theta|do(x)) &= \frac{1}{\sqrt{2\pi}\,\sigma_T} \int dT \; \e{-\frac{T^2}{2\sigma_T^2}} \; \delta\(\theta - x - \alpha\, T\) \nonumber\\
&= \frac{1}{\sqrt{2\pi}\,\sigma_T\, \alpha} \;\e{- \frac{(\theta-x)^2}{2\,\sigma_T^2 \,\alpha^2}} \label{eq:causDep}
\end{align}

While the distributions resulting form the two above setups (Equations~(\ref{eq:statDep}) \mbox{and~(\ref{eq:causDep}))} are clearly distinct, we want to take one further step to show that the information metrics they induce are also different. First, we note that the distributions $ p(y| \theta) $, and~so the ``effect metric'' $ g(\theta) $ (see Equation~(\ref{eq:CG_construct})), simply reflect how the space of decay curves $ y(t) $ is parametrized by the scalar $ \theta $. Thus, in~this example, $ g(\theta) $ is not about causality, and~so is fixed regardless of how we measure our data. We therefore expect this distinction to be captured entirely by the intervention metrics. To~find these, according to the setup in Equation~(\ref{eq:CG_construct}), we first need the Bayesian inverse $ \tilde{q}(do(x)|\theta) $, for~which we must calculate the normalization $ \int dx \; q(\theta|do(x)) $. For~the distribution in Equation~(\ref{eq:causDep}), this is one, and~$\tilde{q}(do(x)|\theta) = q(\theta|do(x))$, thus giving the uniform intervention metric $ h_{caus}(\theta) = \frac{1}{\sigma_T^2 \,\alpha^2} $ for the true causal dependences in the system. For~Equation~(\ref{eq:statDep}), the~integral for normalization cannot be done exactly, so we take the approximation that fluctuations $ \sigma_T $ of the confounding temperature variable $ T $ are small. With~this, we find that $ \int dx \; q(\theta|x) = 1 - \alpha\, \hat{x} \(\frac{\sigma_T}{\sigma_x}\)^2 - \(\frac{3}{2} \alpha\, \hat{x}\, \theta^2 + \frac{1}{4} \theta^4 + \alpha^2\, \sigma_x^2\) \(\frac{\sigma_T}{\sigma_x}\)^4 + \bigO{\sigma_T^6}$, and~thus, the ``intervention'' metric for the statistical dependencies is $ h_{stat}(\theta) = \<\frac{1}{\sigma_{net}^2}\>_x - 3\( \alpha\, \hat{x} + \,\theta^2\) \(\frac{\sigma_T}{\sigma_x}\)^4 + \bigO{\sigma_T^6}$, where the average $ \<\cdot\>_x $ is taken according to the distribution $\tilde{q}(x|\theta)$. Notably, unlike $ h_{caus} $ above, here, $ h_{stat}(\theta) $ varies with $ \theta $, giving a qualitatively different geometry than the true causal geometry for this~example.

As such, we have shown in an explicit setup how the causal relations $ q(\theta|do(x)) $ can differ from statistical ones $ q(\theta|x) $, leading to distinct geometric structures on the model manifold, which may then produce different schemes for model simplification. In~general, while we may not know all the confounding factors nor the effects they can have, we can simply rely on using the $ do $-operator to extract the true causal relations for us. For~this reason, none of the examples presented in this work make any explicit reference to confounding factors, as~these may be assumed to be plentiful and unknown. For~further discussion on the role of confounding factors and the tools of causal calculus that help to work with complex causal graphs, see~\cite{pearl2009causality}.

\section{Deriving Geometric \textit{EI}}

Here, we will derive the expression for $ EI_g $ in Equation~(\ref{eq:EIg}), and~equivalently in Equation~(\ref{eq:dimmer_EI}). We start from the definition of $ EI $ in Equation~(\ref{eq:EIdef}) and use the near-deterministic model approximation discussed in the main text, and~below. In~Section~\ref{app:1D}, we go through the detailed derivation for the 1D case and~then in Section~\ref{app:multD} overview the steps needed to generalize it to higher dimensions. 
%While the 1D result is not in itself so interesting, lacking much of the rich geometric structure of the higher dimensional case, the derivation is presented here in that context purely for notational simplicity. All the following steps, when appropriate adapted, go through unchanged for the multi-dimensional case. 

\subsection{One-Dimensional~Case} \label{app:1D}
As mentioned in the main text, the~expression for $ EI_g $ presented here only approximates the exact $ EI $ when the mappings from interventions $x\in \mathcal{X} $ to parameters $ \theta\in\Theta $ and to effects $y\in \mathcal{Y} $ are both nearly-deterministic. Explicitly, this means that we can express the probability distributions as Gaussians with small variances. Note that the variance can be different at different points and~in the multi-dimensional case may be anisotropic---as long as it remains sufficiently small everywhere (to be clarified later). This way, we can specify the concrete expression for the construct in Equation~(\ref{eq:CG_construct}):
% start a new page without indent 4.6cm

%\end{paracol}
%\nointerlineskip

%\appendixtitles{yes} % Leave argument ''no'' if all appendix headings stay EMPTY (then no dot is printed after ''Appendix A''). If~the appendix sections contain a heading then change the argument to ''yes''.
%\appendix
\renewcommand{\theequation}{A\arabic{equation}}
\setcounter{equation}{2}

%MDPI: same here as for equation/s (6) - should this be one set of equations numbered by (A3) in the middle? Or can each one be numbered individually? %Just single numbering - thanks.

\begin{equation} \label{eq:appCG_gauss}
\begin{split}
&\mbox{interventions }{x}\in\mathcal{X}, \mbox{ parameters } {\theta}\in\Theta \mbox{, effects } {y}\in\mathcal{Y}\\
&{x}\to q({\theta}\,|\,do({x})), \quad \mbox{such that }\;\;
\tilde{q}(do({x})\, |\, {\theta}) \equiv \frac{q({\theta}\, |\, do({x}))}{\int d{x}\; q({\theta}\, |\, do({x}))} = \mathcal{N}_x\(F(\theta), \delta^2\)\\
&{\theta} \to p({y}\,|\, do({\theta}))=\mathcal{N}_y\(f(\theta),\epsilon^2\)\\
&g({\theta}) = \(f'(\theta)/\epsilon\)^2 \quad \mbox{effect metric} \\ 
&h({\theta}) = \(F'(\theta)/\delta\)^2 \quad \mbox{intervention metric}
\end{split}
\end{equation}

%\begin{paracol}{2}
%\linenumbers
%\switchcolumn

Note that in 1D, the~metrics $ g $ and $ h $ become scalars, and~$ \mathcal{N}_x\(F(\theta), \delta^2\) $ denotes a Gaussian distribution in $ x $, centered on $ F(\theta) $ and with standard deviation $ \delta $. Furthermore, we define $ F(\theta) $ and intervention errors $ \delta $ as above, giving that \textls[-35]{$ q({\theta}\,|\,do({x}))=\mathcal{N}_\theta\(F^{-1}(x),\(F^{-1'}(x) \,\delta\)^2\) $}---merely for convenience of notation later (as $ \delta $ may depend on $ x $). As~such, we also assume $ F(\theta) $ and $ f(\theta) $ to be~invertible.

We begin with the definition of $ EI $ from Equation~(\ref{eq:EIdef}), for~which we must first calculate the distribution $ P(y\,|\,do(x)) \equiv \int d\theta \;p(y\,|\, do(\theta))\, q({\theta}\,|\,do({x}))$:
\begin{align} \label{eq:appP0}
P(y\,|\,do(x)) = \int d\theta \;
\overbrace{\frac{1}{\sqrt{2\pi}\, \epsilon} \e{-\frac{(y-f(\theta))^2}{2 \epsilon^2}}}^{p(y\,|\, do(\theta))}\,\times \,
\overbrace{\frac{F'(F^{-1}(x))}{\sqrt{2\pi}\, \delta} \e{-\frac{(x-F(\theta))^2}{2 \delta^2}}}^{q({\theta}\,|\,do({x}))}
\end{align}
We can evaluate this Gaussian integral in the limit of small $ \epsilon $ and $ \delta $. Let us understand precisely how small these must~be. 

To work with the above integral, $ \delta $ must be small enough that both $ \tilde{q}(do(x)\,|\, \theta) $ and $ q(\theta\,|\, do(x)) $ are Gaussian. For~this, the~second-order term in the Taylor expansion $ F(\theta) = F(\theta_x) + (\theta-\theta_x)F'(\theta_x) + \frac{1}{2} (\theta-\theta_x)^2 F''(\theta_x) +...$ around $ \theta_x \equiv F^{-1}(x) $ must be negligible in all regions with substantial probability, giving the asymptotic assumption: $ \frac{1}{2} (\theta-\theta_x)^2 F''(\theta_x) \ll (\theta-\theta_x)F'(\theta_x) $. We can check that in this case, plugging this expansion back into $ \tilde{q}(x \,|\,{\theta}) $ from Equation~(\ref{eq:appCG_gauss}), we get now a Gaussian in $ \theta $: $ \exp{-\frac{(\theta-\theta_x)^2 F'(\theta_x)^2}{2\,\delta^2}} $ as desired. This then shows us that we expect $ \theta $ to typically be within $ \bigO{\delta / F'(\theta_x)} $ of $ \theta_x $, i.e.,~$ (\theta-\theta_x) \sim \bigO{\delta / F'(\theta_x)} $. This allows us to write the above asymptotic condition as $ F''(\theta)\, \delta \ll F'(\theta)^2 $ for all $ \theta$. The~exact same argument goes for the effect distribution $ p(y\,|\, do(\theta)) $ and~similarly gives the condition $ f''(\theta)\, \epsilon \ll f'(\theta)^2 $. As~long as these two conditions hold for all $ \theta $, we can allow $ \delta=\delta(x) $ and $ \epsilon=\epsilon(y) $ to vary~arbitrarily.

In this limit, the~integration of the expression in Equation~(\ref{eq:appP0}) is straightforward to carry out, giving another Gaussian:
%MDPI: same again here: %yes, thanks

\begin{equation} \label{eq:appP}
\begin{split}
&P(y\,|\,do(x)) = \frac{1}{f'(\theta_y) \sqrt{2\pi\, \sigma^2(\theta_x,\theta_y)}} 
\;\e{-\frac{(\theta_x - \theta_y)^2}{2\,\sigma^2(\theta_x,\theta_y)}} \\
\mbox{with}&\qquad \sigma^2(\theta_x,\theta_y)\equiv 
\(\frac{\epsilon}{f'(\theta_y)}\)^2 + \(\frac{\delta}{F'(\theta_x)}\)^2 = 
\frac{1}{g(\theta_y)} + \frac{1}{h(\theta_x)}
\end{split}
\end{equation}
%\begin{align} \label{eq:appP}
%&P(y\,|\,do(x)) = \frac{1}{f'(\theta_y) \sqrt{2\pi\, \sigma^2(\theta_x,\theta_y)}} 
%\;\e{-\frac{(\theta_x - \theta_y)^2}{2\,\sigma^2(\theta_x,\theta_y)}} \nonumber\\
%\mbox{with}&\qquad \sigma^2(\theta_x,\theta_y)\equiv 
%\(\frac{\epsilon}{f'(\theta_y)}\)^2 + \(\frac{\delta}{F'(\theta_x)}\)^2 = 
%\frac{1}{g(\theta_y)} + \frac{1}{h(\theta_x)}
%\end{align}
where $ \theta_y \equiv f^{-1}(y)$ and $ \theta_x \equiv F^{-1}(x)$, and~we used the expressions for the two metrics in Equation~(\ref{eq:appCG_gauss}). Averaging this over the interventions and~using the fact that $ \sigma \sim \bigO{\epsilon,\delta} $ is small, we can then find the effect distribution:
\begin{align}
E_D(y) = \<P(y\,|\,do(x))\>_{I_D(x)} = \int \frac{dx}{L}\; P(y\,|\,do(x))= \frac{F'(\theta_y)}{L\, f'(\theta_y)}
\end{align}
where $ L $ is the size of the 1D intervention space $ \mathcal{X} $, so that the uniform intervention distribution $ I_D(x)=1/L $. 

With these expressions, we can now calculate the $ EI = \<D_{KL}\left [P(y \,|\, do(x)) \; \| \; E_D(y)\right ] \>_{I_D(x)}$. Since $ \sigma $ is small, the~Gaussian for $ P(y\,|\,do(x)) $ will ensure that $ \theta_x $ is close to $ \theta_y $, and~so to leading order, we replace $ \sigma(\theta_x,\theta_y) \approx \sigma(\theta_x,\theta_x) $ here. Therefore:% start a new page without indent 4.6cm

%\end{paracol}
%\nointerlineskip

\begin{align}
EI &= \int \frac{dx}{L} \int \underbrace{\frac{dy}{f'(\theta_y)}}_{=d \theta_y} \;\frac{1}{ \sqrt{2\pi\, \sigma^2(\theta_x)}} 
\;\e{-\frac{(\theta_x - \theta_y)^2}{2\,\sigma^2(\theta_x)}}
\(-\frac{\(\theta_x - \theta_y\)^2}{2 \,\sigma^2(\theta_x)} - \log\left [f'(\theta_y) \sqrt{2\pi\, \sigma^2(\theta_x)}\right ] - \log\left [\frac{F'(\theta_y)}{L\, f'(\theta_y)}\right ]\) 
\nonumber\\
&= \int \frac{dx}{L} 
\(-\frac{1}{2} - \log\left [f'(\theta_x) \sqrt{2\pi\, \sigma^2(\theta_x)}\right ] - \log\left [\frac{F'(\theta_x)}{L\, f'(\theta_x)}\right ]\) \nonumber\\
&= -\frac{1}{2}\; \int \frac{d\theta \,F'(\theta)}{L} \; \log \left [2\pi e \(\frac{F'(\theta)}{L}\)^2\(\(\frac{\epsilon}{f'(\theta)}\)^2 + \(\frac{\delta}{F'(\theta)}\)^2\)\right ] \label{eq:appEI0}\\
&= \log\left [ \frac{L}{\delta \,\sqrt{2\pi e}}\right ] - \frac{\delta}{L} \int d\theta \sqrt{h(\theta)} \;\log\sqrt{h(\theta)\(h^{-1}(\theta) + g^{-1}(\theta)\)} \label{eq:appEI1}
\end{align}
%\begin{paracol}{2}
%\linenumbers
%\switchcolumn

where in the last line, we simply rearranged and substituted the expressions for the metrics $ g(\theta) $ and $ h(\theta) $ from Equation~(\ref{eq:appCG_gauss}). Here, we first see that Line \ref{eq:appEI0} reproduces the $ EI $ approximation we showed for the dimmer switch example, in~Equation~(\ref{eq:dimmer_EI}), for~the setup there: $ F(\theta)=\theta $, and~$ L=1 $. In~general, by~recognizing that here, the volume of the intervention space is $ V_I = \int d\theta\sqrt{h(\theta)} =\int d\theta \,F'(\theta)/\delta = L/\delta $, we finally see that the expression in Equation~(\ref{eq:appEI1}) agrees with our main result presented in Equation~(\ref{eq:EIg}) in the case of a 1D parameter space discussed~here. 

\subsection{Multi-Dimensional~Case} \label{app:multD}
Generalizing the above derivation to the multi-dimensional case is straightforward and~is mainly a matter of careful bookkeeping. The~expressions in Equation~(\ref{eq:appCG_gauss}) become here: 
% start a new page without indent 4.6cm

%\end{paracol}
%\nointerlineskip

%MDPI: same again here %yes. Also: this equation runs into the bottom margin - can force a page break to avoid?

\begin{equation} \label{eq:appCG_multD}
\begin{split}
&\mbox{interventions }{{x}^a}\in\mathcal{X}, \mbox{ parameters } {{\theta}^{\mu}}\in\Theta \mbox{, effects } {{y}^i}\in\mathcal{Y}\\
&\v{x}\to q(\v{\theta}\,|\,do(\v{x})), \quad \mbox{such that }\;\;
\tilde{q}(do(\v{x})\, |\, \v{\theta}) \equiv \frac{q(\v{\theta}\, |\, do(\v{x}))}{\int d^d\v{x}\; q(\v{\theta}\, |\, do(\v{x}))} = \frac{\sqrt{\det {\Delta}_{ab}}}{\(2\pi\)^{d_I/2}} \; \e{-\frac{1}{2}\Delta_{ab}\(x-F({\theta})\)^a \(x-F({\theta})\)^b }\\
&\v{\theta} \to p(\v{y}\,|\, do(\v{\theta}))=\frac{\sqrt{\det E_{ij}}}{\(2\pi\)^{d_E/2}} \; \e{-\frac{1}{2}E_{ij}\(y-f(\theta)\)^i\(y-f(\theta)\)^j}\\
&`g_mn`(\v{\theta}) = E_{ij}\;\partial_\mu f^i(\theta)\, \partial_\nu f^j(\theta)\quad \mbox{effect metric} \\ 
&`h_mn`(\v{\theta}) = \Delta_{ab}\;\partial_\mu F^a(\theta)\, \partial_\nu F^b(\theta) \quad \mbox{intervention metric}
\end{split}
\end{equation}

%\begin{paracol}{2}
%%\linenumbers
%\switchcolumn

Here, $ a,b \in \{1,...,d_I\} $ index the various dimensions of the intervention space $ \mathcal{X} $, $ \mu,\nu \in \{1,...,d\} $----for dimensions of the parameter space $ \Theta $ and~$ i,j \in \{1,...,d_E\} $---the effects space $ \mathcal{Y} $. As~we are dealing with general multi-dimensional geometric constructs, in~this section, we are being careful to denote the contravariant vector components with upper indices, as~in $ \theta^\mu $, and~covariant components with lower indices, as~in $ \partial_\mu \equiv \pd{}{\theta^\mu} $.

As in Equation~(\ref{eq:appP0}), \ref{eq:appP} above, we can then compute the distribution over effects conditioned on interventions:
%MDPI: same again here: %yes, thanks

\begin{equation}
\begin{split}
&P(\v{y}\,|\,do(\v{x})) = \frac{\sqrt{\det {\Sigma}_{\mu\nu}(\theta_x,\theta_y)}}{\det\(\partial_\mu f^i(\theta_y)\) \(
	2\pi\)^{d/2}} 
\;\e{-\frac{1}{2}\Sigma_{\mu\nu}\;(\theta_x - \theta_y)^\mu (\theta_x - \theta_y)^\nu} \\
%\mbox{with}&\qquad \v{\Sigma}^{-1}(\theta_x,\theta_y)\equiv 
%\v{g}^{-1}(\theta_y) + \v{h}^{-1}(\theta_x)
\mbox{with}&\qquad \v{\Sigma}^{-1}(\theta_x,\theta_y)\equiv 
\v{g}^{-1}(\theta_y) + \v{h}^{-1}(\theta_x) \label{eq:appPmult}
\end{split}
\end{equation}
%\begin{align}
%&P(\v{y}\,|\,do(\v{x})) = \frac{\sqrt{\det {\Sigma}_{\mu\nu}(\theta_x,\theta_y)}}{\det\(\partial_\mu f^i(\theta_y)\) \(
%	2\pi\)^{d/2}} 
%\;\e{-\frac{1}{2}\Sigma_{\mu\nu}\;(\theta_x - \theta_y)^\mu (\theta_x - \theta_y)^\nu} \nonumber\\
%%\mbox{with}&\qquad \v{\Sigma}^{-1}(\theta_x,\theta_y)\equiv 
%%\v{g}^{-1}(\theta_y) + \v{h}^{-1}(\theta_x)
%\mbox{with}&\qquad \v{\Sigma}^{-1}(\theta_x,\theta_y)\equiv 
%\v{g}^{-1}(\theta_y) + \v{h}^{-1}(\theta_x) \label{eq:appPmult}
%\end{align}
where $ \v{\Sigma}$ denotes the matrix with components $ \Sigma_{\mu\nu} $ and~$ \v{\Sigma}^{-1} $ its matrix inverse (and similar for $ \v{g} $ and $ \v{h} $). Here, we assume that both functions $ \v{F}(\theta) $ and $ \v{f}(\theta) $ are invertible, which means that the intervention and effect spaces $ \mathcal{X} $ and $ \mathcal{Y} $ both have the same dimension as the parameter space $ \Theta $: $ d_I=d_E=d $. This allows us to view the map $ \theta^\mu \to f^i(\theta) $ as a change of coordinates, with~a square Jacobian matrix $ \partial_\mu f^i $, whose determinant in the first line of Equation~(\ref{eq:appPmult}) is thus well-defined and~may be usefully expressed as $ \det \(\partial_\mu f^i\) = \sqrt{\frac{\det g_{\mu\nu}}{\det E_{ij}}}$. Note also that to get the above result, we once again assume the distributions $ \tilde{q}(do(\v{x})\, |\, \v{\theta}) $ and $ p(\v{y}\,|\, do(\v{\theta})) $ to be nearly deterministic, meaning here that the matrices $ \v{\Delta} $ and $ \v{E} $ must be large, though~the precise form of the assumption is messy here. 
%The assumptions of near-deterministic mappings, discussed in the last section, here takes the form $ f''(\theta)\, \epsilon \ll f'(\theta)^2 $

Averaging this result over the intervention space $ \mathcal{X} $, we get:
\begin{align}
E_D(\v{y})=\<P(\v{y}\,|\,do(\v{x}))\>_{\mathcal{X}} = \frac{\det\(\partial_\mu F^a(\theta_y)\) }{ \det\(\partial_\mu f^i(\theta_y)\)} \frac{\sqrt{\det \Delta_{ab}}}{V_I}
\end{align}
where we define the intervention-space volume in units of variance of $ \tilde{q} $, as: $ V_I \equiv \int d^d\v{x}\sqrt{\det \Delta_{ab}} = \int d^d\v{\theta} \sqrt{\det{h}_{\mu\nu}(\v{\theta})} $. Performing another such average over $ \mathcal{X} $, with~some algebra, similar as for Equation~(\ref{eq:appEI0}), we can arrive at our result in Equation~(\ref{eq:EIg}):
% start a new page without indent 4.6cm

%\end{paracol}
%\nointerlineskip
\begin{align}
EI = \<D_{KL}\left [P(\v{y}\,|\,do(\v{x})) \| E_D(\v{y})\right ]\>_{\mathcal{X}} = 
\log\left [\frac{V_I}{(2\pi e)^{d/2}}\right ]-
\frac{1}{V_I}\int d^d\v{\theta} \sqrt{\det\v{h}} \;\log\, \sqrt{\det \(\ds{1} + \v{g}^{-1}\,\v{h}\)}
\end{align}
%\begin{paracol}{2}
%%\linenumbers
%\switchcolumn

%%%%%%%%%%%%%%%%%%%%%%%%%%%%%%%%%%%%%%%%%%
%\end{paracol}
\reftitle{References}
\bibliographystyle{mdpi}
\renewcommand\bibname{References}

% The following MDPI journals use author-date citation: Arts, Econometrics, Economies, Genealogy, Humanities, IJFS, JRFM, Laws, Religions, Risks, Social Sciences. For those journals, please follow the formatting guidelines on http://www.mdpi.com/authors/references
% To cite two works by the same author:~\citeauthor{ref-journal-1a} (\citeyear{ref-journal-1a},~\citeyear{ref-journal-1b}). This produces: Whittaker (1967, 1975)
% To cite two works by the same author with specific pages:~\citeauthor{ref-journal-3a} (\citeyear{ref-journal-3a}, p. 328;~\citeyear{ref-journal-3b}, p.475). This produces: Wong (1999, p. 328; 2000, p. 475)

%% for journal Sci
%\reviewreports{\\
%Reviewer 1 comments and authors' response\\
%Reviewer 2 comments and authors' response\\
%Reviewer 3 comments and authors' response
%}

%%%%%%%%%%%%%%%%%%%%%%%%%%%%%%%%%%%%%%%%%%
\end{document}